\documentclass{aa}
\usepackage{txfonts}
\usepackage{graphicx}
\usepackage{natbib}
\usepackage{rotating}
\bibpunct{(}{)}{;}{a}{}{,}
\newcommand{\be}{\begin{equation}}
\newcommand{\ee}{\end{equation}}

\newcommand{\obs}[1]{[#1]_{{\tsty {\ssty obs}}}}
\newcommand{\bigobs}[1]{\left[#1\right]_{{\tsty {\ssty obs}}}}
\newcommand{\gen}[1]{#1_{\ssty i}}

\newcommand{\ssty}{\scriptscriptstyle}
\newcommand{\tsty}{\textstyle}

\newcommand{\etal}{et al.\ }

\newcommand{\dl}{d_{\ssty L}}
\newcommand{\da}{d_{\ssty A}}
\newcommand{\dg}{d_{\ssty G}}

\newcommand{\nc}{n_{\ssty C}}
\newcommand{\dz}{d_z}

\newcommand{\Mg}{{\cal M}}

\newcommand{\bibpath}[1]{\bibliography{#1}}
\newcommand{\densplot}[1]{\includegraphics[width=8.5cm]{#1}}
\newcommand{\densplotbig}[1]{\includegraphics[width=17cm]{#1}}

\vspace{+1cm}
\begin{document}
\title{Relativistic cosmology number densities in void-Lema\^\i tre-Tolman-Bondi models}
\titlerunning{Number densities in void cosmologies}
\author{A.~Iribarrem \inst{\ref{OV}} \thanks{iribarrem@astro.ufrj.br}
\and P.~Andreani \inst{\ref{ESO}}
\and S.~February \inst{\ref{UCT}}
\and C.~Gruppioni \inst{\ref{INAF}}
\and A.~R.~Lopes \inst{\ref{OV}}
\and M.~B.~Ribeiro \inst{\ref{IF}}
\and W.~R.~Stoeger \inst{\ref{VO}}
\authorrunning{Iribarrem, Andreani, February \etal}
}
\institute{
Observat\'orio do Valongo, Universidade Federal do Rio de Janeiro, Brazil 
\label{OV}
\and
European Southern Observatory,
Karl-Schwarzschild-Stra\ss e 2, 85748 Garching, Germany
\label{ESO}
\and
Astrophysics, Cosmology and Gravitation Centre,
and Department of Mathematics and Applied Mathematics,
University of Cape Town, Rondebosch 7701, Cape Town, South Africa
\label{UCT}
\and
INAF -- Osservatorio Astronomico di Bologna,
Via Ranzani 1, I-40127 Bologna, Italy
\label{INAF}
\and
Instituto de F\'\i sica, Universidade Federal do Rio de Janeiro, Brazil
\label{IF}
\and
Vatican Observatory Research Group, Steward Observatory,
University of Arizona, AZ 85721, Tucson, USA
\label{VO}
}
\date{ }
\abstract{}
{The goal of this work is to compute the number density of far-IR selected galaxies
in the comoving frame and along the past lightcone of observationally constrained 
Lema\^\i tre-Tolman-Bondi ``giant void'' models and to compare those results with their
standard model counterparts.}
{We derived integral number densities and differential number densities using different cosmological distance definitions in the Lema\^\i tre-Tolman-Bondi dust models. Then, we computed selection functions and consistency functions for the luminosity functions
in the combined fields of the {\em Herschel}/PACS evolutionary probe (PEP) survey in both
standard and void cosmologies, from which we derived the observed values of the above-mentioned densities. We used the Kolmogorov-Smirnov statistics
to study both the evolution of the consistency functions and its connection to the evolution of the comoving density of sources. Finally, we fitted the power-law behaviour of the densities along the observer's past lightcone.}
{The analysis of the comoving number density shows that the increased flexibility of the Lema\^\i tre-Tolman-Bondi models is not enough to fit the observed redshift evolution of the number counts, if it is specialised to a recent best-fit giant void parametrisation. The results for the power-law fits of the densities along the observer's past lightcone show general agreement across both cosmological models studied here around a slope of -2.5 $\pm$ 0.1 for the integral number density on the luminosity-distance volumes. The differential number densities show much bigger slope discrepancies.}
{We conclude that the differential number densities on the observer's past lightcone were still rendered dependent on the cosmological model by the flux limits of the PEP survey. In addition, we show that an intrinsic evolution of the sources must be assumed to fit the comoving number-density redshift evolution in the giant void parametrisation for the Lema\^\i tre-Tolman-Bondi models used in this work.}

\maketitle

\section{Introduction}
\label{intro}
The aim of many studies in observational cosmology is to infer to what extent the geometry of the spacetime
contributes to the formation and evolution of the galaxies, i.e. the building blocks of
the luminous Universe. 
In practical terms, most of what we can infer about galaxy formation and evolution comes
from analyses of redshift surveys.
Although the redshift is an observable quantity related to the energy content
and geometry of the Universe -- {\em regardless of how we model it} -- translating
these measurements into distance estimations cannot be performed without assuming a
cosmological model. As a consequence, it is clear that any study that involves galaxy distances
will be model dependent. This dependency cannot be overcome as long as the distance estimators used 
in these studies are not directly obtained, but instead derived from redshift measurements.

The standard cosmological model, often called the concordance model, is presently
able to simultaneously fit most of the current observations of independent
cosmological quantities. A few examples of observations that support this model
include Hubble parameter measurements from the distances to
Cepheids \citep{2001ApJ...553...47F}, or from the massive, passively evolving early-type
galaxies \citep{2012JCAP...08..006M, 2012arXiv1208.6502}, or from the extragalactic HII
regions \citep{2012MNRAS.425L..56C}; the luminosity distance-redshift relation stemming
from supernovae type Ia (henceforth, simply SNe) surveys
\citep{1998AJ....116.1009R, 1999ApJ...517..565P}, the power spectrum of the cosmic
microwave background radiation \citep[CMB,  e.g.][]{2011ApJS..192...18K},
and the angular size scale obtained from baryonic acoustic oscillation
(BAO) studies \citep[e.g.][]{2010MNRAS.401.2148P}. The degree of confidence on
the concordance model is such that the above-mentioned model dependency of galaxy
formation and evolution nowadays sounds like an unavoidable, but otherwise less
consequential fact.

Despite the sizeable constraining power of the current set of observations
show, there is still room for considering alternative cosmological models.
Even if most of these models are disfavoured when compared to the standard model \citep[e.g. ][]{2009ApJ...703.1374S}, some of them cannot be ruled out entirely yet. 
Among these models, motivated
mainly as alternative to dark energy, various non-homogeneous cosmologies
have been proposed, with an increasing interest in the past few years
\citep{2009PhRvD..79d3501H, 2010GReGr..42.1935A, 2010MNRAS.405.2231F, 2010JCAP...11..030B,
2010CQGra..27q5001S, 2011JCAP...09..011B, 2011PhRvD..83f3506N, 2011JCAP...02..013C,
2012MNRAS.426.1121C, 2012GReGr..44..567R, 2012MNRAS.419.1937M, 
2012GReGr..44.3197H, 2012PhRvD..85j3511N, 2012PhRvD..85j3512B, 2012arXiv1209.4078V,
2012ApJ...748..111W, 2012JCAP...01..043H, 2013ApJ...762L...9H, 2012PhRvD..85b4002B,
2012arXiv1208.4534D, 2012ApJ...754..131K, 2013arXiv1304.7791F}.
Some recent reviews include \citet{2007astro.ph..2416C}, \citet{2011CQGra..28p4002B}, \citet{2011CQGra..28p4001E}, and \citet{2013arXiv1309.4368K}.

The simplest non-homogeneous model assumes a Lema\^{\i}tre-Tolman-Bondi
(LTB) metric coupled to a pressure-less (dust) energy-momentum tensor.
LTB dust models yield analytical solutions to the Einstein's 
field equations, e.g. \citet{1972MNRAS.159..261B}, that can be reduced by an appropriate
parametrization to the standard model ones. It is, for example, quite usual to
set the free equations allowed in this model in a way that it reduces to the usual
Einstein-de Sitter solution at large enough redshifts, enabling the model to
naturally fit many CMB constraints, albeit with very low $H_0$. This model can be understood as a generalisation of the FLRW metric, an effective model, possibly the simplest one,
in which the observed large-scale inhomogeneities play a role in the late-time
dynamics and nearby geometry of the Universe.

Using the greater flexibility of these models,
\citet{1997MNRAS.292..817M} showed that any spherically symmetric astronomical
observation, such as redshift surveys, can be fitted by the non-homogeneities
allowed in these models, instead of any source evolution. This is
particularly relevant for galaxy evolution since it implies that
in order to establish it beyond this theoretical possibility, it is necessary
to use a combination of various independent observable quantities.
\citet[][hereafter GBH]{2008JCAP...04..003G} propose a
``giant void'' parametrisation for the LTB dust model that was able to reach that exactly: fit simultaneously the SNe Ia hubble diagram and some general features of the CMB. Despite the many problems this alternative model shows \citep{2010JCAP...12..021M, 2011PhRvD..83j3515M, 2011PhRvL.107d1301Z, 2011PhRvD..84l3508Z, 2013arXiv1303.5090P}, arguably it remains the simplest and best-studied way to allow for non-homogeneities in the cosmological model. Given the theoretical possibility of fitting the redshift evolution of the number counts without any intrinsic evolution of the sources mentioned above, we aimed to investigate what our view of galaxy evolution would be under these alternative models.

To address the question of how robustly the current observations render the
luminosity function of galaxies, in \citet{I2012b} we used the dataset of FIR-selected sources in \citet{2013MNRAS.tmp.1158G} for the combined GOODS-N, GOODS-S, COSMOS and
ECDF-S fields in the {\em Herschel}/PACS evolutionary probe (PEP) galaxy redshift
survey, computed in both the $\Lambda$CDM standard and LTB giant void models,
the latter parameterized as in \citet{2008JCAP...04..003G}
with the best-fit parameters given by \citet{2012JCAP...10..009Z}.

While the motivation of \citet{I2012b} was to probe the implications of the
underlying cosmology on the redshift evolution of observable properties of the
sources, such as number count and luminosity, the present work aims to further
study these implications, specifically, on the characterization of the number
density of sources along the past lightcone of the assumed cosmology.
The past lightcone is a direct observable: it is the only region of the Universe
manifold that galaxy redshift surveys probe.
In computing the densities in the lightcone, we remove the model assumption
intrinsic to the usual comoving frame computations. The empirical approach of
it, if not as ambitious, shares the same philosophy of the {\em ideal observational
cosmology} programme as \citet{1985PhR...124..315E}, where the aim was to determine the
spacetime geometry of the Universe without assuming a cosmological model {\em a priori}.

The present line of work started with \citet{2003ApJ...592....1R}, which connected the
relativistic cosmology number counts results summarized in \citet{1971grc..conf..104E}
to the practical luminosity function (LF) results from galaxy redshift surveys.
The LF is a statistical tool to infer the formation and evolution of galaxies from a set of photometric or spectroscopic data. To compute the luminosities of the sources from their observed
fluxes, a luminosity distance must be obtained from the redshift estimation,
a step that requires adopting a cosmological model. To deal with incompleteness due to the flux limits, an assumption
on the {\em spatial homogeneity} of the distribution is usually made. Spatial homogeneity is defined on constant time-coordinate hypersurfaces, not to be confused with {\em observational homogeneity} that is
defined on the observer's past lightcone. The spatial homogeneity assumption
does not hold for non-homogeneous cosmologies in general, but \citet{I2012b}
showed that it does not lead to significant differences in the LF results
in the case of the LTB/GBH models adopted here.

\citet{2005A&A...429...65R} considers the effect of the expansion of the spacetime in the number densities along the past lightcone leading to observational non-homogeneities in the spatially homogeneous Friedmann-Lema\^{\i}tre-Robertson-Walker (FLRW) model. \citet{2008A&A...488...55R} expanded on those theoretical results,
showing that homogeneous distributions in the past lightcone would lead to spatial
non-homogeneity, in disagreement with the usual form of the {\em cosmological principle}.

\citet{2007ApJ...657..760A} combined the results of the first two previous papers
to compute such lightcone distributions using the luminosity functions for the
CNOC2 survey from \citet{1999ApJ...518..533L}, confirming the presence of such light
cone non-homogeneities. In \citet{2012A&A...539A.112I} we
discussed further ways to compare the relativistic effect of the expanding spacetime
to the evolution of the sources in the FLRW past lightcone, using the much wider
redshift range in the luminosity functions for the FORS Deep Field survey from
\citet{2004A&A...421...41G, 2006A&A...448..101G}.

The goal of the present work is to include non-FLRW spacetimes in the past lightcone
studies described above. We specialised the general equations of \citet{2003ApJ...592....1R}
to the LTB metric, combining the results from both
\citet{1992ApJ...388....1R}, and \citet{2008JCAP...04..003G}. Finally, we used the luminosity
functions in \citet{I2012b}, which were computed by assuming the LTB/GBH model from their
build-up.

     The question we aim to address is the following: are the
number densities along the past lightcone and the characterisation
of their power law behaviour, robust among the standard
and void cosmological models cited above? In addition, we discuss the comoving number density in the void models,
with possible implications for empirical models like that in, e.g.,
\citet{2011MNRAS.416...70G, 2012arXiv1208.6512B}.

The paper is organised as follows.
In Sect. \ref{theory}, we specialise the results in \citet{2003ApJ...592....1R}
and \citet{2005A&A...429...65R} to the constrained GBH parametrisation of
the LTB dust models. In Sect. \ref{sfsection}, we compute the selection functions
for the FIR datasets in \citet{I2012b}. In Sect. \ref{Jsection} we compute the
connection between the predicted comoving density of sources in both FLRW
and LTB geometries to the selection functions. In Sect. \ref{gammasection} we
compute the number densities of PEP sources down the past lightcone of
both cosmologies. In Sect. \ref{results}, we present the results, show evidence of galaxy evolution in the LTB/GBH model, and compare the comoving number density evolution over both cosmologies. In Sect. \ref{discussion} we discuss the dependency of the relativistic number densities on both flux limit and cosmological model. In Sect. \ref{conclusions} we present our conclusions.

\section{Theoretical quantities in the LTB/GBH model}
\label{theory}

In the empirical approach of \citet{2007ApJ...657..760A}, the number
densities along the observer's past lightcone were computed from LF
data using different relativistic cosmology distance definitions. These
distances are defined in the same cosmological model as
the one assumed on the build-up of the LF. In this context no assumptions on the redshift evolution of the sources is made. The methodology is completely empirical.

The number densities used in this paper, as defined in
\citet{2005A&A...429...65R}, are able to probe the geometrical effect
of the expansion of Universe on the homogeneity of the distribution of the
sources along the observer's past lightcone. This effect depends on
the distance definition used in computing these densities, which in turn
depends on the line element of the cosmological model assumed.

In this section we connect the key results from \citet[][henceforth R92]{1992ApJ...388....1R}
for the number count of sources in the LTB metric to the
parametrisation advanced by \citet{2008JCAP...04..003G} for that cosmology.
The goal here is to compute the above-mentioned number densities in the
giant void parametrisation of the LTB model. Dotted quantities refer
to time-coordinate derivatives and primed ones refer to radial-coordinate
derivatives.

\subsection{Differential number count}
\label{dNdzthsec}
We started by writing the line element for the LTB model following
\citet{1972MNRAS.159..261B}
\begin{equation}
\label{LTBmetric}
ds^2 = -c^2 dt^2 + \frac{A'^2(r,t)}{f^2(r)} dr^2 + A^2(r,t) d\Omega ^2,
\end{equation}
where $d\Omega ^2 = d\theta ^2 + \sin ^2 \theta \, d \phi ^2$,
with $f(r)$ and $A(r,t)$ being arbitrary functions.
Assuming a pressure-less
(dust) matter content with $\rho_{\ssty M}$ proper density, it can be
shown that the Einstein's field equations for that line element can be
combined to yield (R92)
\begin{equation}
\label{rho}
8 \pi G \rho_{\ssty M} = \frac{F'}{2 A' A^2},
\end{equation}
where $F(r)$ is another arbitrary function satisfying the relation
above, and $G$ is the gravitational constant.
Starting from the general expression for the number count of sources
derived by \citet{1971grc..conf..104E}, R92 obtained
\begin{equation}
\label{R92eq26}
dN = 4 \pi \, n \, \frac{A' A^2}{f} \, dr,
\end{equation}
where $n$ is the number density per unit proper volume, and $N$ the
total number of sources down the past lightcone. Assuming as an order of
magnitude estimation of the average total mass of each source
$\Mg \approx 10^{11} {\cal M}_{\ssty \sun}$, R92 wrote
\begin{equation}
\label{R92eq27}
n = \frac{F'}{16 \pi  G \, \Mg \, A' A^2}.
\end{equation}
Combining the last two equations, we obtained
\begin{equation}
\label{dNdr}
\frac{dN}{dr} = \frac{1}{4G \, \Mg} \, \frac{F'}{f}.
\end{equation}

The last equation is essentially a version of the result of \citet{1971grc..conf..104E},
specialised to the past null geodesic of the LTB model. Next, we further specialised it to use the
GBH parametrisation for their constrained model, and the best-fit
values obtained by \citet{2012JCAP...10..009Z} in a simultaneous
analysis of SNe Ia, CMB, and BAO data.

It is straightforward to relate $f(r)$ to the spatial curvature
parameter k(r) in GBH by writing
\begin{equation}
\label{fk}
f(r) = \sqrt{1-k(r)}.
\end{equation}
The boundary condition equations listed in GBH read as
\begin{eqnarray}
H_{\ssty T}(r,t) = \frac{\dot{A}(r,t)}{A(r,t)}  \label{GBHHT}, \\
F(r) = 2 \, \Omega_{\ssty M}(r) \, H^2_{\ssty 0}(r) \, r^3 \label{GBHF}, \\
k(r) = -[1-\Omega_{\ssty M}(r)] \, H^2_{\ssty 0}(r) \, r^2 \label{GBHk},
\end{eqnarray}
with the gauge choice $A(r,0)=r$ included, where $H_{\ssty T}$
is the transverse Hubble rate, $H_{\ssty 0}(r) = H_{\ssty T}(r,0)$,
and $\Omega_{\ssty M}(r)$ is the dimensionless matter density parameter.
This last quantity is defined
relative to the integrated critical density in the comoving volume
at radial coordinate $r$ as
\begin{equation}
\label{zumarhoc}
\bar{\rho}_{\ssty C} = \frac{3 \, H^2_{\ssty 0}(r)}{8 \pi \, G}.
\end{equation}

The present-time transverse
Hubble parameter, $H_{\ssty 0}(r)$, in the constrained version of the GBH
model is parametrised as
\begin{equation}
\label{zumaH0}
H_{\ssty 0}(r) = H_{\ssty in} \sum_{n=0}^{\infty}
\frac{2 \, [1-\Omega_{\ssty M}(r)]^n}{(2n+1)(2n+3)}.
\end{equation}
Equations (\ref{fk}), (\ref{GBHk}), and (\ref{zumaH0}) can be readily combined to yield
\begin{equation}
\label{kr}
f(r) = \sqrt{1 + [1-\Omega_{\ssty M}(r)] \left\{ \sum_{n=0}^{\infty}
\frac{2 \, [1-\Omega_{\ssty M}(r)]^n}{(2n+1)(2n+3)}
\, \right\}^2 {H_{\ssty in}}^2 r^2}.
\end{equation}

Combining Eqs. (\ref{GBHF}) and (\ref{zumaH0}), we can write
the $F(r)$ function for the constrained model as
\begin{equation}
\label{Fr}
F(r) = H_{\ssty in} \left\{ \sum_{n=0}^{\infty}
\frac{2 \, [1-\Omega_{\ssty M}(r)]^n}{(2n+1)(2n+3)}
\, \right\} \Omega_{\ssty M}(r) \, r^3,
\end{equation}
where the dimensionless matter density parameter
$\Omega_{\ssty M}(r)$ in the GBH model becomes
\begin{equation}
\label{omr}
\Omega_{\ssty M}(r) = \Omega_{\ssty out} + (\Omega_{\ssty in} -
\Omega_{\ssty out}) \left\{ \frac{1-\tanh[(r-R)/2\Delta r]}
{1+\tanh[R/2\Delta r]} \right\}.
\end{equation}
Here $H_{\ssty in}$ is the transverse Hubble constant at the center of the
void, $\Omega_{\ssty in}$ is the density parameter at the center
of the void, $\Omega_{\ssty out}$ the asymptotic density parameter
at large comoving distances, $R$ is the size of the under-dense
region, and $\Delta R$ the width of the transition between the
central void and the exterior homogeneous region. 
These parameters completely determine the model.

Because of the generality of the evolution of 
$A(r,t)$ in the LTB models, the time coordinate $t_{\ssty bb}$ at which $A(r,t_{\ssty bb})$
reduces to zero, can, in general, assume different values for different comoving
distances from the centre of the under-dense region,
$r$. This leads to different measurements for the elapsed time since the
Big Bang, $t_{\ssty bb}(r)$, depending on the position of the observer in the void.
Setting this extra degree of freedom for the big-bang time in order to make it
simultaneous (same value for all observers) yields the constrained version of the
GBH model, or CGBH model. The best-fit values we use in this work were obtained
in \citet{2012JCAP...10..009Z}, considering both an asymptotically flat CGBH model
with $\Omega_{\ssty out} = 1$ and an open CGBH model (OCGBH) with $\Omega_{\ssty out} = 0.87$, which the authors show better fits the CMB constraints.
Table \ref{zumafit} reproduces these values.

\begin{table*}
\caption[Best-fit values for the LTB/CGBH models.]{Best fit values for the LTB/CGBH models from \citet{2012JCAP...10..009Z} used in this work.
\label{zumafit}}
\begin{tabular}{lll}
\hline
\hline
Model parameter				&CGBH						&OCGBH				 	\\
\hline
$H_{\ssty in}$				&66.0 $\pm$ 1.4					&71.1 $\pm$ 2.8				\\
$\Omega_{\ssty in}$ 		&0.22 $\pm$ 0.4					&0.22 $\pm$ 0.4				\\
$R$ [Gpc] 					&0.18 $^{+0.64}_{-0.18}$		&0.20 $^{+0.87}_{-0.19}$	\\
$\Delta R$ [Gpc]			&2.56 $^{+0.28}_{-0.24}$		&1.33 $^{+0.36}_{-0.32}$	\\
$\Omega_{\ssty out}$ 		&1								&0.86 $\pm$ 0.03			\\
\hline
\end{tabular}
\end{table*}

For each comoving distance $r$, $\Omega_{\ssty M}(r)$,
$f(r)$, and $F(r)$ can be computed through Eqs. (\ref{kr}), and (\ref{Fr}).
The radial derivative $F'(r)$ can be obtained
numerically through central difference quotients,
$F(r) \approx \lim_{\Delta r \ll r} \Delta F / \Delta r$,
which in turn allow the values of $dN/dr$ to be computed through Eq.
(\ref{dNdr}).

\begin{figure}
\centering
\densplot{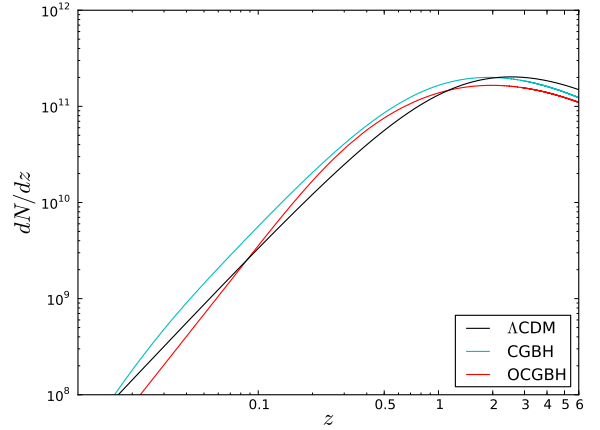}
\caption{Differential number count estimates within the past lightcone of the three cosmological models used in the present work. 
%
\label{plotdNdz}}
\end{figure}

The redshift can be related to the radial coordinate in this
model following \citet{2007JCAP...02..019E}
\begin{equation}
\label{drdz}
\frac{dr}{dz} = \frac{1}{1+z} \, \frac{f}{\dot{A}'},
\end{equation}
and to the time coordinate through the incoming radial null
geodesic equation as in GBH
\begin{equation}
\label{dtdz}
\frac{dt}{dz} = - \frac{A'}{f(r)}.
\end{equation}
Following GBH, all the first- and second-order derivatives of $A(r,t)$ can be
written as power series expansion, which
allows Eqs. (\ref{drdz}) and (\ref{dtdz}) to be solved
numerically. This yielded the look-back time and radial distance tables: $t(z)$ and $r(z)$.
We then combined Eqs. (\ref{dNdr}) and (\ref{drdz}) as
\begin{equation}
\label{dNdz}
\frac{dN}{dz} = \frac{1}{4G \, \Mg} \, \frac{F'[r(z)]}{(1+z) \, \dot{A}'[r(z),t(z)]},
\end{equation}
and computed the differential number count $dN/dz$ for each value
of $z$ in the r(z) and t(z) tables. A comparison between the estimates for this
quantity in the $\Lambda$CDM and both CGBH parametrisations in
\citet{2012JCAP...10..009Z} can be found in Fig. \ref{plotdNdz}.

\subsection{Distances and number densities}
\label{gammathsec}

Since it is impossible to measure the distance to galaxies directly,
there are a few different ways of deriving this quantity from other
measurements. The luminosity distance, for example, is based on the relation
between the emitted and received fluxes of the source, whereas the angular
diameter distance is based on the relation between the observed angular size
and the actual physical size of the source. Those relations, however, depend
on the expansion history of the Universe and will, in general, lead to
different results simply because their dependency on the underlying
cosmological model is not the same.

The angular diameter distance in LTB models can be identified
as $\da(z) = A[r(z),t(z)]$. From that we obtain the luminosity distance
$\dl(z)$ and the galaxy-area distance $\dg$ through Etherington's reciprocity
law \citep{1933PMag...15..761E, 1971grc..conf..104E, 2007GReGr..39.1047E}
\begin{equation}
\label{reclaw}
\dl = (1+z)^2 \da = (1+z) \dg,
\end{equation}
whereas their redshift derivatives, $d(\da)/dz$, $d(\dl)/dz$, and $d(\dg)/dz$ 
can be computed numerically with the same method used for computing $F'(r)$.
To compare the theoretical results for the number densities in the
past lightcone of the LTB/GBH models with those for the FLRW geometry given
in \citet[Figs. 5-6]{2007ApJ...657..760A}, we also compute the
{\em redshift distance} $\dz$ simply as $\dz = c z / H_{\ssty 0}$.

With that we can compute the relativistic {\em differential number densities} ($\gen{\gamma}$), the number of sources per unit volume in a spherical shell at redshift $z$, for each distance $\gen{d}$ in $i=[A, G, L, z]$ and in each cosmological model as in \citet{2005A&A...429...65R}:
\begin{equation}
\label{gammath}
\gen{\gamma} = \frac{dN}{dz} \left\{ 4 \pi \, (\gen{d})^2 \,
\frac{d(\gen{d})}{dz} \right\}^{-1}.
\end{equation}

Following \citet{2012A&A...539A.112I}, the {\em integral number densities} ($\gen{\gamma^{\ssty \ast}}$), i.e. the number of sources per unit volume located inside the observer's past lightcone down to redshift $z$, can be computed for each distance definition $\gen{d}$ as
\begin{equation}
\label{gstarth}
\gen{\gamma^{\ssty \ast}} = \frac{3 \, N}{4\pi \, (\gen{d})^3},
\end{equation}
with the cumulative number count $N(z)$ obtained simply by
\begin{equation}
\label{Nth}
N(z) = \int_0^{z} \frac{dN}{dz}(z') \, dz'.
\end{equation}

\begin{figure}
\centering
\densplot{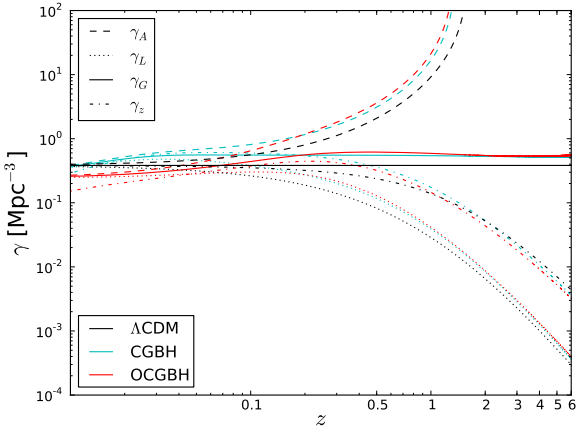}
\caption{Redshift evolution of the relativistic differential densities for the three cosmological
models used in the present work. Different curves correspond to the computations performed with
respect to different distance estimators along the observer's past lightcone ($\da$, $\dl$, $\dg$,
and $dz$).
%
\label{plotgmth}}
\end{figure}

\begin{figure}
\centering
\densplot{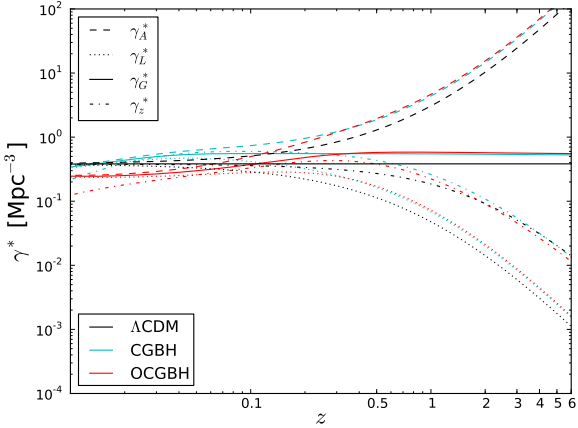}
\caption{Integral of the functions shown in Fig. \ref{plotgmth}. The curves show the evolution
of the relativistic integral densities for the three cosmological
models used in the present work as a function of the different distances.
\label{plotgsth}}
\end{figure}

Results for the various $\gen{\gamma}$ and $\gen{\gamma^{\ssty \ast}}$
in the different cosmologies considered in this work are plotted in
figures \ref{plotgmth} and \ref{plotgsth}. We note that
among these cosmological models, the differential and integral
number densities show noticeable differences in all distance definitions used, even if apparently small to the eye. These differences
may actually be observed, depending on the precision achieved by a galaxy survey.
We go on to check whether this purely geometrical effect can be
detected on the LF for the PEP survey computed in \citet{I2012b}.
This requires considering the luminosity and the number evolution
of the galaxies in the survey volume.

\section{Selection functions}
\label{sfsection}

Selection functions are an estimate of the number density of
galaxies with luminosity above a chosen threshold, L$_{\ssty lim}$. Once
computed, the selection functions can be
used to estimate the observed number of objects per comoving volume,
derived in Sect. \ref{gammathsec}, and from that the observed differential
and integral densities in the past lightcone.

Following the empirical approach of \citet{2012A&A...539A.112I},
we computed selection functions $\psi$ in each redshift
interval using each of the three cosmological models ($\Lambda$CDM, CGBH, and OCGBH) and the monochromatic 100 $\mu$m and 160 $\mu$m rest-frame luminosity functions for the combined PEP fields in \citet{I2012b} as
\begin{equation}
\label{sfeq}
\psi_{\ssty \bar{z}} = \int_{L_{\ssty lim,\,\bar{z}}}^
{L_{\ssty max,\,\bar{z}}} \phi_{\ssty \bar{z}}(L) \, dL \, ,
\end{equation}
where $L_{\ssty max,\,\bar{z}}$ is the brightest source luminosity inside the redshift interval, and
$L_{\ssty lim,\,\bar{z}}$ is the rest-frame luminosity associated
to the flux limit of the observations.

The dataset used in this work was built using combined fields, i.e observations carried out 
in different sky regions at different depths. The large number of sources per redshift interval allowed
us to derive the observed quantities with better statistics, at the expense of an added difficulty
related to the definition of the luminosity limits corresponding to the actual
flux limits of the observations. This difficulty stems from the fact
that the computation of the rest-frame luminosity involves
k corrections, which in turn depend on the spectral energy
distribution (SED) of the source. That is, each SED template defines
a slightly different luminosity limit for the same flux limit.
In the case of the PEP survey, each field -- namely GOODS-N, GOODS-S,
COSMOS and ECDF-S -- also had a different flux limit, depending on
the PACS passband in which the observation was done
\citep{2011A&A...532A..90L}.

To investigate how important the selection function variations
caused by the different SED templates in the datasets are, we first
computed the selection functions in each redshift interval assuming
the lowest computed luminosity among the sources in that interval,
and then compared the results to the averages and to the highest luminosities.
The variations in the selection functions caused by assuming
different luminosity limits were found to be many times greater than
the error bars propagated from the luminosity function uncertainties.
For example, at $z$ = 0.2, the selection function computed using the
lowest monochromatic luminosity in the 100-$\mu$m dataset read
(9.0 $\pm$ 1.8) $\times$ 10$^{-3}$ Mpc$^{-3}$, while the one computed using the
average luminosity read (3.7 $\pm$ 0.7) $\times$ 10$^{-3}$ Mpc$^{-3}$, and the
one using the highest luminosity read (2.6 $\pm$ 0.5) $\times$ 10$^{-3}$ Mpc$^{-3}$ -
a variation more than three times larger than the combined uncertainties obtained
by propagating the ones from the LF parameters. For the same dataset,
at $z$ = 1, the selection functions read as (2.3 $\pm$ 0.3) $\times$ 10$^{-3}$ Mpc$^{-3}$,
(1.1 $\pm$ 0.1) $\times$ 10$^{-3}$ Mpc$^{-3}$, and (4.9 $\pm$ 0.6) $\times$ 10$^{-4}$ Mpc$^{-3}$,
respectively, a variation almost six times larger than the propagated
uncertainties.

Therefore, for every source in each redshift interval,
we computed the rest-frame luminosity for the flux limit of the field
and filter where it was observed, given its best-fit SED and redshift.
Next, we computed a set of selection function values in that
redshift interval, using the luminosity limits computed above for each
source in the interval.
Finally, we computed the average over this set of selection function values
for a given redshift interval and used this average as the value for the selection function in that same interval. The uncertainties, as discussed above, are dominated by the variation in the luminosity limits and therefore can be taken simply as the standard deviation over the same set of computed selection function values for each redshift interval. The resulting monochromatic
rest-frame 100 $\mu$m and 160 $\mu$m selection functions are given in Tables
\ref{psi100table} and \ref{psi160table}.

%

\begin{table} 
{\tiny 
\caption[Selection functions for the rest-frame 100 $\mu$m datasets.]{Selection functions for the rest-frame 100 $\mu$m datasets. Units are Mpc$^{-3}$.\label{psi100table}} 
\centering 
\begin{tabular}{llll} 
\hline \hline 
\multicolumn{1}{l}{$\bar{z}$}&\multicolumn{1}{c}{$\psi^{\ssty \Lambda CDM}$} &\multicolumn{1}{c}{$\psi^{\ssty CGBH}$} &\multicolumn{1}{c}{$\psi^{\ssty OCGBH}$}  \\ \hline 
  0.1 &$(4.0 \pm 0.8)  \times 10^{-3}$     &$(5.0 \pm 1.0)  \times 10^{-3}$     &$(5.6 \pm 1.1)  \times 10^{-3}$     \\ 
  0.3 &$(3.6 \pm 0.5)  \times 10^{-3}$     &$(4.4 \pm 0.6)  \times 10^{-3}$     &$(5.0 \pm 0.7)  \times 10^{-3}$     \\ 
  0.5 &$(2.6 \pm 0.4)  \times 10^{-3}$     &$(3.2 \pm 0.5)  \times 10^{-3}$     &$(3.6 \pm 0.5)  \times 10^{-3}$     \\ 
  0.7 &$(2.3 \pm 0.4)  \times 10^{-3}$     &$(2.9 \pm 0.5)  \times 10^{-3}$     &$(3.2 \pm 0.6)  \times 10^{-3}$     \\ 
  0.9 &$(1.3 \pm 0.2)  \times 10^{-3}$     &$(1.9 \pm 0.3)  \times 10^{-3}$     &$(1.9 \pm 0.3)  \times 10^{-3}$     \\ 
  1.1 &$(1.1 \pm 0.3)  \times 10^{-3}$     &$(1.6 \pm 0.4)  \times 10^{-3}$     &$(1.6 \pm 0.4)  \times 10^{-3}$     \\ 
  1.4 &$(4.3 \pm 1.1)  \times 10^{-4}$     &$(3.7 \pm 1.0)  \times 10^{-4}$     &$(4.8 \pm 1.2)  \times 10^{-4}$     \\ 
  1.6 &$(4.1 \pm 1.4)  \times 10^{-4}$     &$(3.6 \pm 1.2)  \times 10^{-4}$     &$(4.6 \pm 1.5)  \times 10^{-4}$     \\ 
  2.0 &$(5.1 \pm 1.6)  \times 10^{-4}$     &$(5.9 \pm 1.9)  \times 10^{-4}$     &$(6.0 \pm 1.9)  \times 10^{-4}$     \\ 
  2.4 &$(4.9 \pm 1.7)  \times 10^{-4}$     &$(5.6 \pm 2.0)  \times 10^{-4}$     &$(5.7 \pm 2.0)  \times 10^{-4}$     \\ 
  2.8 &$(3.3 \pm 1.1)  \times 10^{-4}$     &$(1.7 \pm 0.5)  \times 10^{-4}$     &$(2.5 \pm 0.8)  \times 10^{-4}$     \\ 
  3.2 &$(2.7 \pm 1.1)  \times 10^{-4}$     &$(1.4 \pm 0.6)  \times 10^{-4}$     &$(2.1 \pm 0.8)  \times 10^{-4}$    \\ \hline 
\end{tabular}} 
\end{table}
\begin{table} 
{\tiny 
\caption[Selection functions for the rest-frame 160 $\mu$m datasets.]{Selection functions for the rest-frame 160 $\mu$m datasets. Units are Mpc$^{-3}$.\label{psi160table}} 
\centering 
\begin{tabular}{llll} 
\hline \hline 
\multicolumn{1}{l}{$\bar{z}$}&\multicolumn{1}{c}{$\psi^{\ssty \Lambda CDM}$} &\multicolumn{1}{c}{$\psi^{\ssty CGBH}$} &\multicolumn{1}{c}{$\psi^{\ssty OCGBH}$}  \\ \hline 
  0.1 &$(6.0 \pm 1.4)  \times 10^{-3}$     &$(3.7 \pm 0.9)  \times 10^{-3}$     &$(3.7 \pm 0.9)  \times 10^{-3}$     \\ 
  0.3 &$(5.4 \pm 0.9)  \times 10^{-3}$     &$(3.3 \pm 0.5)  \times 10^{-3}$     &$(3.3 \pm 0.5)  \times 10^{-3}$     \\ 
  0.5 &$(2.7 \pm 0.5)  \times 10^{-3}$     &$(2.6 \pm 0.4)  \times 10^{-3}$     &$(3.0 \pm 0.5)  \times 10^{-3}$     \\ 
  0.7 &$(2.5 \pm 0.5)  \times 10^{-3}$     &$(2.3 \pm 0.5)  \times 10^{-3}$     &$(2.7 \pm 0.6)  \times 10^{-3}$     \\ 
  0.9 &$(1.0 \pm 0.2)  \times 10^{-3}$     &$(1.3 \pm 0.2)  \times 10^{-3}$     &$(1.8 \pm 0.3)  \times 10^{-3}$     \\ 
  1.1 &$(9.8 \pm 2.7)  \times 10^{-4}$     &$(1.2 \pm 0.3)  \times 10^{-3}$     &$(1.8 \pm 0.4)  \times 10^{-3}$     \\ 
  1.4 &$(4.6 \pm 1.2)  \times 10^{-4}$     &$(4.4 \pm 1.2)  \times 10^{-4}$     &$(4.2 \pm 1.1)  \times 10^{-4}$     \\ 
  1.6 &$(4.5 \pm 1.3)  \times 10^{-4}$     &$(4.3 \pm 1.3)  \times 10^{-4}$     &$(4.1 \pm 1.2)  \times 10^{-4}$     \\ 
  2.0 &$(1.1 \pm 0.3)  \times 10^{-4}$     &$(3.0 \pm 0.8)  \times 10^{-4}$     &$(2.4 \pm 0.7)  \times 10^{-4}$     \\ 
  2.4 &$(1.3 \pm 0.3)  \times 10^{-4}$     &$(3.4 \pm 0.9)  \times 10^{-4}$     &$(2.8 \pm 0.8)  \times 10^{-4}$     \\ 
  2.8 &$(5.9 \pm 1.9)  \times 10^{-5}$     &$(1.5 \pm 0.4)  \times 10^{-4}$     &$(1.4 \pm 0.4)  \times 10^{-4}$     \\ 
  3.2 &$(3.5 \pm 1.6)  \times 10^{-5}$     &$(9.2 \pm 4.3)  \times 10^{-5}$     &$(8.6 \pm 4.0)  \times 10^{-5}$    \\ \hline 
\end{tabular}} 
\end{table}

\section{Consistency functions}
\label{Jsection}

In the empirical framework of 
\citet{2003ApJ...592....1R}, \citet{2007ApJ...657..760A}, and \citet{2012A&A...539A.112I},
the observed quantities computed in the past lightcone of a given
cosmological model were obtained from their predicted values through
what was called a completeness function $J(z)$. This can be somewhat
confusing from the observer's point of view, since $J(z)$ is not
necessarily related to incompleteness like that of missing sources in
a survey, but to the relation between a theoretical prediction for the
number counts and the actual measurement of that quantity
\citep{2012A&A...539A.112I}.
In this sense, this quantity would be better named as {\em consistency function},
a term we use from now on. The consistency function $J(z)$
was obtained by relating the prediction for the comoving number density $\nc$
given by the cosmological model
\begin{equation}
\label{ncth}
\nc(z) = \frac{\rho_{\ssty M} (z)}{\Mg} = \frac{\Omega_{\ssty M}[r(z)]}{\Mg} \, \rho_{\ssty C},
\end{equation}
to the selection functions for a given galaxy survey $\psi_{\ssty \bar{z}}$
in a given redshift interval $\bar{z}$ as \citet{2012A&A...539A.112I}
\begin{equation}
\label{Jcalc}
J_{\ssty \bar{z}} = \frac{\psi_{\ssty \bar{z}}}{\nc(\bar{z})}.
\end{equation}
The values for $J_{\ssty \bar{z}}$ can be obtained numerically for the CGBH
void models using Eqs. (\ref{zumarhoc}) and (\ref{omr}) combined with
the appropriate $r(z)$ table as described in \S\ref{theory} and the selection
functions computed in \S\ref{sfsection}.


As discussed in \citet{2012A&A...539A.112I}, the estimation of the
comoving number density $\nc$ assumes a constant, average mass value $\Mg$
for normalisation purposes -- that is, getting a correct order-of-magnitude
estimation of the number density. Such estimation is not supposed to
provide a detailed description of $\nc(z)$, but rather to
convey the information on the redshift evolution of the comoving number density
caused solely by the cosmological model, through $\rho_{\ssty M} (z)$, in
Eq. (\ref{ncth}). The details of the redshift evolution of
the total masses of the sources, missing in this estimation, are imprinted on
the observed LF and inherited by its derived selection functions. By
translating this purely theoretical $\nc(\bar{z})$ estimation to the corresponding
selection functions $\psi_{\ssty \bar{z}}$ through the consistency function
$J(\bar{z})$, we allow any theoretical quantity that assumes a $\nc(z)$ built on a cosmological model to use the observed values given by the selection functions.

For the purpose of obtaining the relativistic number densities empirically,
this approach is sufficient. It minimizes the number of theoretical assumptions
about the evolution of the sources, such as including a \citet{1974ApJ...187..425P}
formalism, and considers as much information as possible from the observations.
Since a Press-Schechter-like formalism is still not implemented on LTB models,
the empirical approach described above is the simplest way to work with both standard and void cosmologies in a consistent way.



\section{Observed number densities in the past lightcone}
\label{gammasection}

We computed the relativistic number densities using the $J_{\ssty \bar{z}}$
obtained in section \ref{Jsection}.
By definition, the LF is the number of sources per unit luminosity, per unit
comoving volume. This allows us to identify the selection functions in Eq. (\ref{sfeq})
as the differential comoving density of galaxies stemming from the observations
and to rewrite Eq. (\ref{Jcalc}) as
\begin{equation}
\label{psi_to_nc}
\psi_{\bar{z}} = \bigobs{\frac{dN}{dV_{\ssty c}}}(\bar{z}) = J_{\bar{z}} \, \nc(\bar{z}) =
J_{\bar{z}} \, \frac{dN}{dV_{\ssty c}}(\bar{z}),
\end{equation}
which can still be written as
\begin{equation}
\label{dNdz_obs}
\bigobs{\frac{dN}{dz}} = J_{\bar{z}} \, \frac{dN}{dz},
\end{equation}
since the purely geometrical term $dV_{\ssty c}/dz$ cancels out on both sides
of the equation. Together with Eqs. (\ref{Jcalc}) and (\ref{dNdz}),
Eq. (\ref{dNdz_obs}) allows us to obtain the differential number counts
$\obs{{dN}/{dz}}$ from the selection function of a given dataset in the
different comoving densities $\nc(z)$ defined by the void model parametrisations
considered here.

Because the quantity inside the parentheses in Eq. (\ref{gammath})
is also a purely geometrical term, we can simply write
\begin{equation}
\label{gammaobs}
\obs{\gen{\gamma}}(\bar{z}) = J_{\bar{z}} \, \gen{\gamma}(\bar{z}),
\end{equation}
which allows us to obtain the observed differential densities $\obs{\gen{\gamma}}$
in the lightcones of the cosmologies considered here once $J_{\bar{z}}$ is computed.

To obtain the observed integral densities $\obs{\gen{\gamma^\ast}}(\bar{z})$,
we substitute $dN/dz$ in Eq. (\ref{Nth}) with $\obs{dN/dz}$ from Eq. (\ref{dNdz_obs}) and
compute the observed cumulative number counts
$\obs{N}(\bar{z})$. Then we can substitute this result back into Eq. (\ref{gstarth}),
since the cosmological distance to redshift relation $\gen{d}(z)$ is also fixed
by the geometry of the given cosmology.

\section{Results}
\label{results}

With the quantities obtained in the last two sections, we can proceed to investigate the differences in the number densities on the comoving frame and on the past lightcone of the standard and void cosmologies.

\subsection{Comoving number density evolution}
\label{Jsub}
As mentioned above, \citet{1997MNRAS.292..817M} showed that any spherically symmetric observation alone, e.g. redshift estimations or number counts, can be fit purely by a general enough LTB dust model, with no evolution of the sources required.
In the context of the present discussion their argument can be understood by  combining Eqs. (\ref{ncth}) and (\ref{Jcalc}) to write
\begin{equation}
\label{Jalt}
J_{\ssty \bar{z}} \, = \, \psi_{\ssty \bar{z}} \, \frac{\Mg}{\rho_{\ssty C} \, \Omega_{\ssty M}(\bar{z})}.
\end{equation}
Looking at the right-hand side of the equation above, one can easily separate its two terms into an observed quantity to the left and a fraction between theoretically obtained quantities to the right. Moreover, one can identify the constant $\Mg$ with the lack of a model for the secular evolution of the average mass of the sources, and $\Omega_{\ssty M}(\bar{z})$
with the evolution of the matter density parameter on a given cosmology.
By using the extra degree of freedom in setting
the matter density profile in the LTB model, $\Omega_{\ssty M}(z)$, one can obtain a number density $\nc(z)$ that matches the selection functions perfectly, without the need to assume an evolving average mass $\Mg (z)$. To constrain this degree of freedom, it is necessary to have different sets of independent observations. This is precisely what the GBH parametrisation of a giant void in an LTB dust model yields: a matter density profile that is parametrised to best fit the combination of different, independent observations.

\begin{figure}
\centering
\densplot{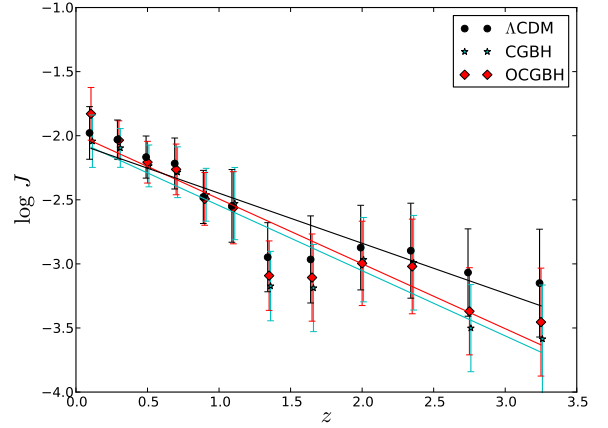}
\caption{Consistency functions for the monochromatic 100 $\mu$m luminosity
functions computed in the three cosmological models used in the present work. These functions are related to the
redshift evolution of the baryon to total mass fraction (see text for detail).
\label{plotJ100}}
\end{figure}

\begin{figure}
\centering
\densplot{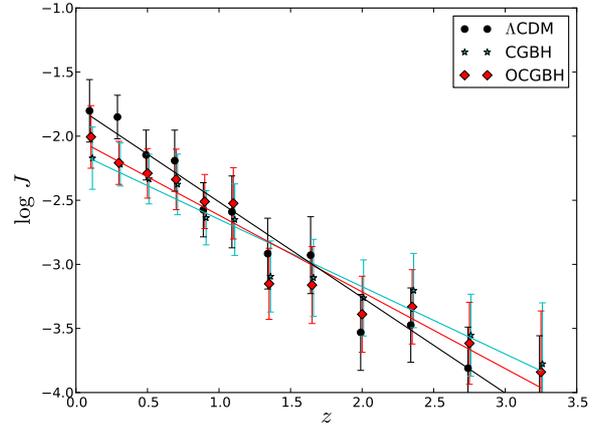}
\caption{As Fig. \ref{plotJ100} for the monochromatic 160 $\mu$m luminosity
functions.
From these plots, one can see that the difference between the standard model consistency
function (black circles) and the void model ones (colour circles) is not very significant,
since most of the points lie within the error bars. The $p_{\ssty \Lambda}$ values
discussed in \S \ref{Jsub} quantify this conclusion and agree with it.
\label{plotJ160}}
\end{figure}

We present the evolution of the consistency functions for the different PACS filters and cosmological models in Figs. \ref{plotJ100} and \ref{plotJ160}. It is clear from these plots that this quantity evolves with redshift in both standard and void cosmologies. The behaviour of the consistency functions in all models studied here are well fit by a power law decreasing with redshift. The best-fitting slopes are given in Table \ref{etable}.
These results suggest that not considering the evolution of the sources
leads to a systematic trend that increases the inconsistency with $z$. This is not as obvious as it may seem given the flexibility of the LTB models as discussed above. It is the parametrisation of $\Omega_{\ssty M}(z)$ constrained by the combined
independent observations of SNe + CMB + BAO that requires $\Mg$ to evolve with $z$ in the LTB
models studied here. Only by allowing $\Mg$ to evolve with $z$ can Eq. (\ref{Jalt})
yield a constant consistency function that indicates that the theoretical term on the
right of the right-hand side of Eq. (\ref{Jalt}) is proportional to the observed term on the left.

We used a Kolmogorov-Smirnoff (KS) two-sample test to check, on a statistical
sense, how different the consistency functions are when assuming the different cosmological models studied here
from a hypothetically constant consistency function. By design, a constant
$J(z)$ indicates that the model for the comoving number density used here, assuming $\Mg(z) = \Mg$,
matches the observed number density. The KS test is a non-parametric, distribution-free way to compare two datasets, and it assigns to which level of confidence we can refute the hypothesis that they were obtained from the same underlying distribution. The resulting p-value of this test can be understood as the probability that both datasets come from the same distribution. For all cosmologies studied
here, the no-evolution hypothesis, $\Mg(z) = \Mg$, is rejected at over 5-$\sigma$
confidence level, with p-values lower than 10$^{-5}$. This means that an evolving
average mass of the sources is also required by the LTB/GBH models studied here.

Next, we use the same KS test to determine how different the consistency functions in the void models are from their standard model counterparts. We present the computed p-values of these tests, $p_{\ssty \Lambda}$, in Table \ref{etable}.
The $p_{\ssty \Lambda}$ value is the probability that the consistency function for
a given void model and the one for the standard model are computed from the same
underlying distribution or, in other words, that their differences are not statistically significant.
To establish whether such a difference is significant, at a 3-$\sigma$ confidence level,
for example, the $p_{\ssty \Lambda}$ value should be lower than approximately 0.0027.
Given the listed $p_{\ssty \Lambda}$ results, we found that the computed consistency functions in void and standard cosmologies were not significantly different.
This means that galaxy evolution proceeds mostly in the same way,
regardless of the differences in the geometry and the composition of the Universe
studied here.

\begin{table} 
{\tiny 
\caption{Comoving, differential and integral densities statistics.\label{etable}} 
\centering 
\begin{tabular}{llllll} 
\hline \hline 
\multicolumn{1}{l}{Dataset} &\multicolumn{1}{l}{Model} &\multicolumn{1}{l}{p$_{\ssty \Lambda}$} &\multicolumn{1}{l}{$\log J(z)$ slope}&\multicolumn{1}{c}{$\log \gamma_{\ssty L}$ slope} &\multicolumn{1}{c}{$\log \gamma^\ast_{\ssty L}$ slope}\\ \hline 
										&$\Lambda$CDM	&1.0                             &-0.39 $\pm$ 0.05                &-2.4 $\pm$ 0.2                  &-2.31 $\pm$ 0.03                 \\ 
L$_{100 \, \mu m}$	&CGBH	&0.43                            &-0.51 $\pm$ 0.06                &-3.1 $\pm$ 0.6                  &-2.50 $\pm$ 0.04                 \\ 
										&OCGBH	&0.43                            &-0.51 $\pm$ 0.05                &-3.0 $\pm$ 0.4                  &-2.52 $\pm$ 0.03                 \\ 
\multicolumn{6}{c}{}  \\ 
										&$\Lambda$CDM	&1.0                             &-0.75 $\pm$ 0.03                &-4.5 $\pm$ 0.3                  &-2.54 $\pm$ 0.06                 \\ 
L$_{160 \, \mu m}$	&CGBH	&0.79                            &-0.53 $\pm$ 0.03                &-3.7 $\pm$ 0.4                  &-2.48 $\pm$ 0.05                 \\ 
										&OCGBH	&0.79                            &-0.60 $\pm$ 0.04                &-3.7 $\pm$ 0.3                  &-2.59 $\pm$ 0.06                 \\ \hline 
\end{tabular}} 
\end{table}


\subsection{Lightcone inhomogeneities}
\label{powerlaw}

For this section, we used the observed differential and integral densities obtained in Sect. \ref{gammasection} to investigate the effects of the central underdensity prescribed by the GBH models in the redshift distortions caused by the expanding spacetime on the observer's past lightcone.

Our previous studies of these relativistic number densities \citep{2005A&A...429...65R, 2007ApJ...657..760A, 2012A&A...539A.112I} suggested a high-redshift power-law behaviour for both differential and integral radial distributions. In other words, that expansion distributes the sources along the lightcone in an increasingly self-similar manner. This is a purely geometrical effect, as discussed in Sect. \ref{theory}, that may or may not be dominant because the hierarchical build-up of galaxies also plays a role in how the sources are distributed along the lightcone. 
Given that some of the alternative void models are built assuming an LTB line element instead of the standard model FLRW one, we aim in this section to better characterise those power laws and to investigate how they are affected by the secular evolution of the sources and by the direct effect of the expansion of
the different geometries.
We focus the discussion on the densities
computed using the luminosity distance, since the results for the other distance
definitions are all qualitatively the same.

As can be seen in Figs. \ref{plotgmth} and \ref{plotgsth}, the geometrical effect increases with the redshift, deviating both the theoretical differential and the cumulative number densities away from a constant, homogeneous behaviour at high redshifts. At redshifts lower than z $\approx$ 0.1, this geometrical effect is less pronounced, and both $\gamma_{\ssty L}(z)$ and $\gamma^\ast_{\ssty L}(z)$ shown follow the constant average galactic mass $\cal M$ assumed in their computation, as discussed in Sect. \ref{sfsection}.

It follows that there must be a region in redshift space at which a transition occurs between the power-law behaviour induced by the expansion at high redshift and the behaviour defined by the underlying density parameter $\Omega_{\ssty M}(z)$ combined with the evolution of the sources stemming from the LF. However, the exact size and range of this region cannot be predicted with the empirical approach for the comoving number densities used in this work. Therefore, if we are to characterise the power-law behaviour of the high-redshift end of our number densities, we must allow our fitting procedure to search for the best redshift at which the power-law behaviour begins to dominate.

\begin{figure*}
\centering
\densplotbig{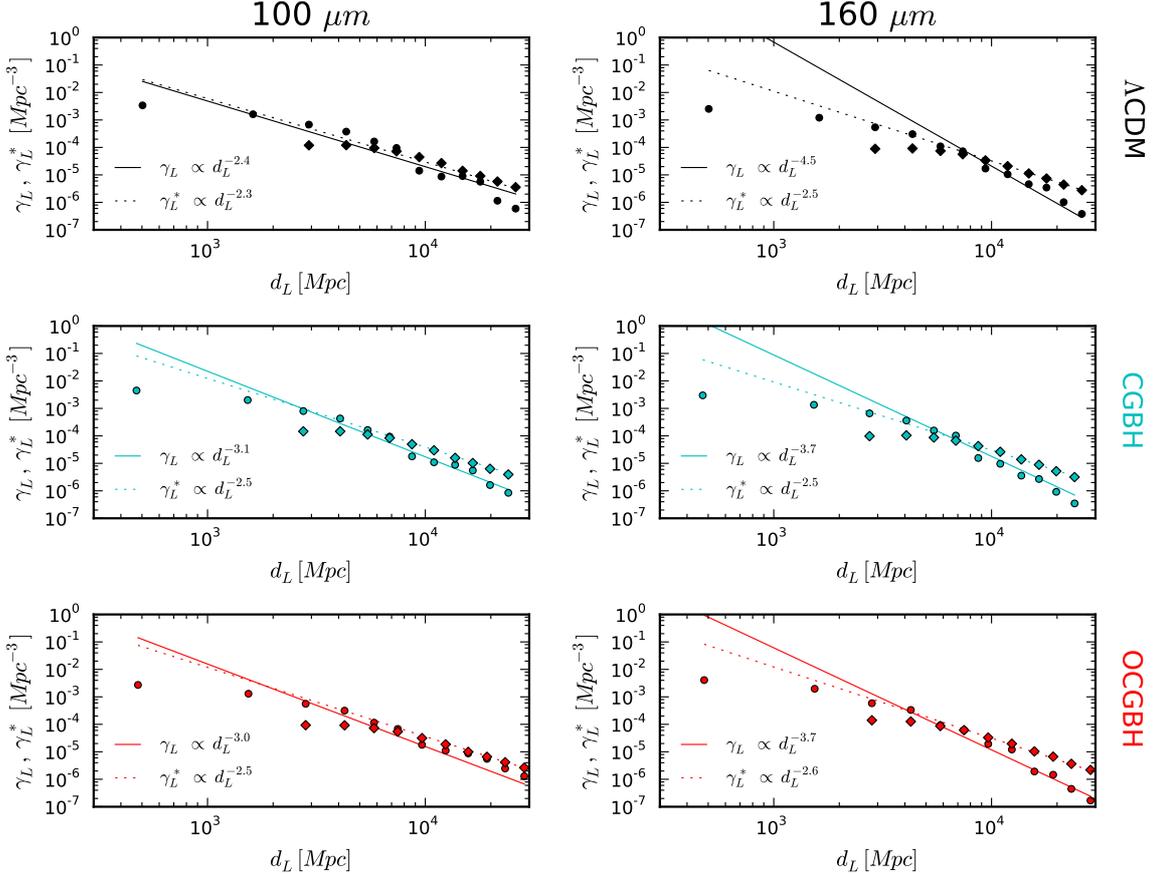}
\caption{Best-fit power laws to the differential density $\gamma_{\ssty L}$ (dots), and
the integral density $\gamma^\ast_{\ssty L}$ (diamonds). The left panels show the plots for the
100 $\mu m$ LF, while the right ones show the plots for the 160 $\mu m$ LF, in the
three cosmological models / parametrisations studied here. Full lines show the best-fit high-redshift power law for the differential densities, while the dashed lines show the best-fit power laws for the integral densities, as discussed in Sect. \ref{powerlaw}.
\label{gvsdlplot}}
\end{figure*}

To do so, we compute the best linear fit to the $\log \gamma_{\ssty L}$ vs. $\log \dl$ tables in an iterative way. 
Starting with the LF values derived for the highest three redshift bins, we perform the same fit, including one extra LF value in each iteration, in decreasing order, until we have included all points. Then, we select the fit with
the lowest reduced $\chi^2$ value. The selected fits for the different dataset and cosmology combinations are plotted in figure \ref{gvsdlplot}. The listed uncertainties are formally obtained taking the square root of the linear term in the covariance matrix of the fitted power law.

To quantify how significant the differences in the slopes of the power laws are, we can start by first computing the uncertainty of the difference $\Delta_{\ssty a,b}$, between the $\alpha_{\ssty a}$ slope of $\log \gamma_{\ssty L}$ using a given 
dataset/cosmology combination $a$ and that of a different combination $b$ as
\begin{equation}
\delta \Delta_{\ssty a,b} \, = \, \sqrt{(\delta \alpha_{\ssty a})^2 + (\delta \alpha_{\ssty b})^2}.
\end{equation}
The significance of this difference in terms of its uncertainty is then obtained
from $\Delta_{\ssty a,b}/(\delta \Delta_{\ssty a,b})$.


\section{Discussion}
\label{discussion}

Next we use the approach above to obtain comparisons between standard vs. void cosmologies, 100 $\mu$m vs. 160 $\mu$m datasets, and differential densities $\gamma$ vs. integral densities $\gamma^\ast$.

\subsection{Comparison with the LTB/CGBH models}
Results for the significance of the difference between slopes of the high-redshift power-law fits to the differential densities $\gamma$ and the integral densities $\gamma^\ast$, given a fixed cosmological model, were
highly dependent on the dataset. They were fairly insignificant with the 100 $\mu$m data for all cosmologies ($\Lambda$CDM: 0.3-$\sigma$, CGBH: 0.9-$\sigma$, OCGBH: 1.1-$\sigma$);
whereas they were all at least marginally significant in the 160 $\mu$m data
($\Lambda$CDM: 5.2-$\sigma$, CGBH: 3.0-$\sigma$, OCGBH: 3.4-$\sigma$).

Such differences in the slopes between the high-redshift power-law fits to $\gamma$ and to $\gamma^\ast$ can be understood by checking Eqs. (\ref{gammath}) and (\ref{gstarth}). We note that $\gamma$ is proportional to $dN/dz$, whereas $\gamma^\ast$ is proportional to $N$. Since $N$ is a cumulative quantity, it can only increase or remain constant with increasing redshift. That is, even if there are regions in the volume where $N$ is not defined,
which have a lower density of detected sources, $N$ itself will remain constant.
On the other hand, $dN/dz$ is sensitive to such local density variations, which adds up to a higher degree of non-homogeneity in the higher redshift part of the past lightcone probed by the survey and as a consequence a steeper power-law slope of $\gamma$.

The differences between the slopes of the high-redshift power-law fits to the relativistic densities computed using the 100 $\mu$m, and the 160 $\mu$m PEP blind-selected datasets were fairly insignificant when assuming the void models studied here (CGBH: 0.7-$\sigma$ for $\gamma$, and 0.3 for $\gamma^\ast$; OCGBH: 1.3-$\sigma$ for $\gamma$, and 1.0 for
$\gamma^\ast$), whereas, those differences showed a moderate-to-strong significance if the $\Lambda$CDM model was assumed (4.6-$\sigma$ for $\gamma$, and 3.0-$\sigma$ for $\gamma^\ast$).

The differences in the slopes obtained using different datasets can be understood as the effect of different redshift-evolving luminosity limits on the past lightcone distributions. The results above indicated that the distributions on the past lightcone of the void models were significantly less different, hence less affected by the detection limits of the PEP datasets, than their counterparts computed on the past lightcone of the standard model.

Finally, we computed the significance of the differences obtained by comparing the same slopes computed in two different cosmological models. The only marginally significant ($\geq$ 3-$\sigma$) difference we found is for the comparison between standard and void models for the integral density $\gamma^\ast$ slopes, using the 100 $\mu$m dataset:
3.7-$\sigma$ for the $\Lambda$CDM vs. CGBH comparison, and 
4.3-$\sigma$ for the $\Lambda$CDM vs. OCGBH one. Given the striking concordance of the other $\gamma^\ast$ slopes around a tentative value of -2.5 $\pm$ 0.1 it is possible that the oddly low values of the slope of $\gamma^\ast$ in the 100 $\mu$m dataset and its uncertainty are an artefact caused by our fitting procedure.

How do we combine the results from all of those different comparisons in a coherent picture?
\citet{I2012b} showed that the corrections needed to build the LF using a flux-limited survey made the results sensitive to the differences caused by changing the underlying cosmological model. Detection limits seem to play a major role in the fully-relativistic analysis used here as well.
On one hand, we found significant differences caused by the kind of
statistics used ($\gamma$ or $\gamma^\ast$) with the 160 $\mu$m blind-selected
catalogue. On the other hand, we found that the cosmology assumed (FLRW/$\Lambda$CDM or LTB/GBH) caused significant differences on both $\gamma$ and $\gamma^\ast$, using the 100 $\mu$m dataset instead. Some of these discrepancies could be caused by the way we fit the power laws to the high-redshift parts of the distributions. Future observations with lower flux limits will help us check which of these discrepancies were caused by the present flux limits.

\subsection{Comparison with other LTB models}

It is important to notice that the results in both Sects. \ref{Jsub} and
\ref{powerlaw}, together with the discussion presented above, are only valid for a very
special case of LTB models, namely, the parametrisation for the CGBH model obtained
by \citet{2012JCAP...10..009Z}. It is not in the scope of the present work to present a
complete analysis of other LTB models in the literature, which would require first
recomputing the luminosity functions presented in \citet{I2012b} from the start,
but given some recent advancements in better exploiting the flexibility of these models,
a qualitative discussion of their possible impact on our results is pertinent here.

Assuming different non-homogeneous cosmologies can affect the present analyses in three
different ways. First, in terms of building the LF from the observations, a model with a matter
density profile $\Omega_M(r)$ that is different enough from the ones studied here could possibly make
the average homogeneity assumption at the heart of the 1/V$_{max}$ LF estimator invalid.
Using mock catalogues, \citet{I2012b} showed that the void shapes of the CGBH models
used in their LF does not significantly affect the ability of the 1/V$_{max}$ estimator
to recover the underlying LF.

Second, different distance-to-redshift relations could also affect the LF results, and thus
its redshift evolution, through the computation of the maximum observable redshift
of each source z$_{max}$, which is used in computing of the 1/V$_{max}$ value of
the LF in each redshift bin. As discussed in \citet{I2012b}, this comes as a result of
the fainter part of the galaxies in a survey having fluxes near the limit of observation
in the field and is affected mainly by differences in the luminosity distance-to-redshift
relation. Also, at luminosities L $\approx$ L$^\ast$, the z$_{max}$ of the sources is
safely hitting the higher $z$ limit of the redshift bin, leading their volume correction to be
independent of their luminosity, hence of the $\dl(z)$ relation.

Finally, the expected number densities in the past-lightcone $\gen{\gamma}$ and
$\gen{\gamma^{\ssty \ast}}$, as computed in Sect. \ref{gammathsec} are
both sensitive to the distance-to-redshift relations, as shown in Figs. \ref{plotgmth}
and \ref{plotgsth}. The differences caused on the number densities by the different distance
relations might be large enough to be detected, given the observational uncertainties
stemming from the LF. Considering the present results, such differences in the expected
number densities should be greater than the CGBH-to-FLRW ones studied here, if we were to
detect a significant effect on the observed number densities caused by differences in their expected values
alone.

\citet{2010MNRAS.405.2231F} argue that CMB and BAO constraints may be significantly
distorted by differences in the evolution of primordial perturbations caused by the
curvature inside the voids. They go on to fit a number of different shapes for the
inner matter profile, considering only local Universe data, namely, SNe Ia and the
reconstruction of $H(z)$ through passively evolving galaxies. They show that all their
models mimic the FLRW $\dl(z)$ to sub-percent level, which should lead to the changes
in the maximum redshift estimates and in the expected number densities caused by assuming
such models comparable to the ones we observe here.

The authors show in their Fig. 9 that
up to redshift z $\approx$ 1, their best-fit models follow closely the distance
modulus, and thus the $\dl(z)$ relation of $\Lambda$CDM. Looking at the same plot,
we can see that the luminosity distances grow increasingly smaller when compared to their
standard model counterparts at higher redshifts. Looking at Eq. (\ref{gstarth}), we can
expect that this could potentially lead to higher values of $\gamma^\ast$ and,
consequentially, a lower value for its best-fit power law.

Also, their best-fit voids are much larger than the ones used here, as can be seen
by comparing the upper left-hand panel of their Fig. 7 versus the upper panel of Fig. 1
in \citet{I2012b}, which shows the best-fit voids of \citet{2012JCAP...10..009Z}.
Mock catalogue tests should be used to make sure these larger voids do not bias the
1/V$_{\ssty max}$ method used here. However, as argued in \citet{I2012b}, if the
large under-dense profiles used here varied smoothly enough not to significantly
bias the LF estimator, we do not expect that the larger voids of \citet{2010MNRAS.405.2231F}
will.

\citet{2011JCAP...02..013C} discuss the implications of allowing for inhomogeneity in the
early time radiation field and the way it affects the constraining power of CMB results for
inhomogeneous models. They show, for example, that even models that are asymptotically
flat at the CMB can be made to fit if the matter density profile includes a low-density
shell around the central void. An under-dense shell like this could possibly lead the
1/V$_{max}$ method to over-estimate the volume correction of sources with maximum observable
comoving volume inside or beyond the shell, because such corrections are based on the
assumption of an average homogeneous distribution. To test in detail how accurately this LF estimator would
recover the underlying distribution, we would need to build mock catalogues
that assume the matter distribution of these models, but we expect that such differences,
if present, would show more prominently at mid to high redshift, according to the size
of the shell, and make the faint-end slope steeper by assigning higher over-estimated corrections
to the fainter sources. Depending on how non-homogeneous the matter profile is, an over-estimation
of the characteristic number density parameter $\phi^\ast$ might also be detected.

Varying bang-time functions $t_{\ssty bb}(r)$ were also studied \citep{2009JCAP...07..029C}
and shown to be an extra degree of freedom that helps LTB models to reconciliate
CMB + H$_0$ constraints. \citet{2012PhRvD..85b4002B}, however, show that models
with varying $t_{\ssty bb}(r)$ constrained by the combination of SNe + CMB + H$_0$
should produce a kinetic Sunyaev-Zel'dovich effect that is orders of magnitude stronger
than its expected upper limits. It is not clear how allowing for this extra degree of
freedom would change the key quantities shown to affect our results: the $\dl(z)$ and
$r(z)$ relations and the shape of the central under-dense region. Only a full-fledged analysis,
starting from the building up of the LF and assuming a model with a varying bang-time fit
by the observations, would allow us to make any reasonable comparison with the results
presented in this work.

\section{Conclusions}
\label{conclusions}

In this paper we derived the theoretical results needed to compute the differential and integral densities along the past lightcone of LTB dust models. We applied
these results to computing the theoretical predictions for such quantities in
the CGBH parametrisations of \citet{2012JCAP...10..009Z}.

We computed the selection functions stemming from the far-IR LFs of
\citet{I2012b}, which allowed us to establish, at an over 5-$\sigma$ confidence
level, that geometry alone is not able to fit their behaviour, given the
LTB/CGBH parametrisation of \citet{2012JCAP...10..009Z}. This finding
confirms the need to allow for evolution of the sources (either in number or in
luminosity) in this particular class of void models as well. We found no strong evidence of any dependence of this evolution on the cosmological models studied here. In other words, the combined merger tree and barionic processes needed to reproduce the redshift evolution of the FIR LF in the CGBH void models are not significantly different from the hierarchical build up and astrophysical processes in the standard model.

Finally, we computed the observed differential and integral densities in the past lightcone of both standard and void cosmologies, and fitted their high-redshift {\em observational non-homogeneities} using power laws. We show that the systematic dependency of the LF methodology on the cosmology that was discussed in \citet{I2012b} could still lead to significant differences in these relativistic number densities.
The integral densities showed a somewhat consistent slope across all combinations of blind-selected datasets and cosmological models studied here. On the other hand, the differential densities were found to be sensitive to a change in cosmological model assumed in their computation. These results confirmed the power-law behaviour of the galaxy distribution on the observer's past lightcone of the LTB/GBH models and allowed a tentative value of -2.5 $\pm$ 0.1 to be obtained for the cumulative radial statistics $\gen{\gamma^{\ssty \ast}}$ of this distribution, regardless of the cosmological model assumed.


\section*{Aknowledgements}
This work was jointly supported by Brazil's CAPES and ESO studentships. SF is supported by the South African Square Kilometre Array (SKA) Project.




\bibliographystyle{aa}
\bibpath{general}
\end{document}